\documentclass[fleqn,usenatbib]{mnras}
\usepackage{amsmath}
\usepackage{graphicx}
\usepackage{grffile}
\usepackage{hyperref}
\usepackage{txfonts}
\usepackage{natbib}
\usepackage{bm}

\title[NGC~2617 in a high state]{The curtain remains open: NGC~2617 continues in a high state}

\author[V. L. Oknyansky et al.]{V. L. Oknyansky,$^{1}$\thanks{E-mail: oknyan@mail.ru}
C. M. Gaskell,$^{2}$
N. A. Huseynov,$^{3}$
V. M. Lipunov,$^{1}$
N. I. Shatsky,$^{1}$
\newauthor S. S. Tsygankov,$^{4}$
E. S. Gorbovskoy,$^{1}$
Kh. M. Mikailov,$^{3}$
A. M. Tatarnikov,$^{1}$
\newauthor D. A. H. Buckley,$^{5}$
V. G. Metlov,$^{1}$
 A. E. Nadzhip,$^{1}$
A. S. Kuznetsov,$^{1}$
P. V. Balanutza,$^{1}$
\newauthor M. A. Burlak,$^{1}$
G. A. Galazutdinov,$^{6}$
 B. P. Artamonov,$^{1}$
I. R. Salmanov,$^{3}$
\newauthor K. L. Malanchev,$^{1}$
R. S. Oknyansky$^{7}$\\
\\
$^{1}$Sternberg Astronomical Institute, M. V. Lomonosov Moscow State University, 119234, Moscow, Universitetsky pr-t, 13, Russian Federation\\
$^{2}$Department of Astronomy and Astrophysics, University of California, Santa Cruz, CA 95064, USA\\
$^{3}$Shamakhy Astrophysical Observatory, National Academy of Sciences, Pirkuli, AZ 5626, Azerbaijan\\
$^{4}$Tuorla Observatory, Department of Physics and Astronomy, University of Turku, V\"ais\"al\"antie 20, FI-21500, Piikki\"o, Finland\\
$^{5}$The South African Astronomical Observatory, PO Box 9, Observatory 7935, South Africa\\
$^{6}$Instituto de Astronomia, Universidad Catolica del Norte, Av. Angamos 0610, Antofagasta, 1270709, Chile\\
$^{7}$Ben Gurion University of Negev, PO Box 653, 8410501,
 Beer-Sheva, Israel}

\date{Accepted XXX. Received YYY; in original form ZZZ}

\pubyear{2016}

\begin{document}
\date{Received ... Accepted ...}
\pagerange{\pageref{firstpage}--\pageref{lastpage}}
\maketitle{}

\label{firstpage}

\begin{abstract}
Optical and near-infrared photometry, optical spectroscopy, and soft X-ray and UV monitoring of the changing look active galactic nucleus NGC~2617 show that it continues to have the appearance of a type-1 Seyfert galaxy.  An optical light curve for 2010--2016 indicates that the change of type probably occurred between 2010 October  and 2012 February  and was not related to the brightening in 2013. In 2016 NGC~2617 brightened again to a level of activity close to that in 2013 April.  We find variations in all passbands and in both the intensities and profiles of the broad Balmer lines.  A new displaced emission peak has appeared in H$\betaup$. X-ray variations are well correlated with UV--optical variability and possibly lead by $\sim$ 2--3 d. The $K$ band lags the $J$ band by about 21.5 $\pm$ 2.5 d. and lags the combined $B+J$ filters by $\sim$ 25 d. $J$ lags $B$ by about 3 d. This could be because
$J$-band variability arises from the outer part of the accretion disc, while $K$-band variability comes from thermal re-emission by dust.  We propose that spectral--type changes are a result of increasing central luminosity causing sublimation of the innermost dust in the hollow biconical outflow.  We briefly discuss various other possible reasons that might explain the dramatic changes in NGC~2617.
\end{abstract}




\section{Introduction}

Active galactic nuclei (AGNs) can be classified on the basis of their optical spectra into `type 1' AGNs (Sy1), showing prominent broad Balmer lines, and  `type 2' AGNs (Sy2), lacking obvious broad Balmer lines \citep{Khachikian+Weedman71}.  Designations such as `Seyfert 1.8' are used for intermediate cases (see \citealt{Osterbrock81}). Rare cases of so-called `changing-look' AGNs (CL AGNs) -- AGNs that show extreme changes of spectral type -- provide  important tests of theories of the Sy1/Sy2 dichotomy.  The first detailed investigations of changes from type 2 to type 1 and back to type 2 were for Mrk~6 \citep{pronik_chuvaev1972,chuvaev1991} and for NGC~4151 \citep{lyutyj1984,penston_perez1984,oknyanskii1991}.  Recent reviews by \citet{shappee2014} and \citet{koay2016} give lists of objects and references.

Some time between 2003 and 2013 the type 1.8 Seyfert NGC~2617 underwent a dramatic change to appear like a Seyfert 1 galaxy \citep{shappee2013,shappee2014}. Before 2013 optical spectra of the object were obtained only twice: in 1994 \citep{moran1996} and in 2003 (the 6dFGS spectrum). The SDSS $ubgriz$ data for NGC~2617 in 2006 showed a very low level of brightness. \citet{shappee2013,shappee2014} therefore suspected that NGC~2617 had been in a low state and had a spectral type close to type 2 for the decade before that but then changed its appearance to type 1, presumably in 2013 April.

CL AGNs such as NGC~2617 are rare. There are currently only some tens of cases known.  However, the small number of known CL AGNs is comparable to the number of AGNs that have had many years of spectral monitoring. It is therefore reasonable to suspect,  that perhaps each strongly variable AGN could be found to be a CL AGN if observed long enough. This assumption is supported by recent results of \citet{runco16} that about 38\% of 102 Seyferts changed type and about 3\% of the objects have disappearing H$\betaup$ on time--scales of $3 - 9$ yr.  Also \cite{macleod16} estimate that $> 15$\% of strongly variable luminous quasars exhibit a changing-look behaviour on rest-frame times--scales of $3000 - 4000$~d.

\citet{Keel80} discovered that orientation plays a key role in whether an AGN is seen as type 1 or type 2.  This led to the simplest unified model of AGN activity (the `straw-person model' or `SPM' of \cite{Antonucci93}), in which all thermal AGNs have a broad-line region (BLR), but in type-2 AGNs we cannot readily see it because our view is blocked by obscuring gas and dust perpendicular to the axis of rotational symmetry.  Despite the popularity of the SPM, there has always been a discussion on the role of {\em other} factors (see \citealt{Antonucci93,Antonucci12}).  CL AGNs provide a challenge to the simplest unified models and would seem to require either that the obscuring dust is very clumpy or that it is being sublimated as the luminosity increases and reforming as the luminosity falls. The dramatic change of Seyfert type and the outbursts observed in NGC~2617 can thus give us valuable information for understanding the geometry, physical nature and evolution of AGNs.

\section{Observations}

We commenced the spectroscopic and photometric monitoring of NGC~2617 in 2016 January  to see if it still appeared to be a type-1 AGN three years after the intensive 2013 monitoring campaign of \cite{shappee2014}. Our observations included IR ($JHK$) and optical ($BVR_cI_c$) photometry and spectroscopy. Additionally, we have used unfiltered optical monitoring by the MASTER robotic network from 2010 to 2016. We found that the nucleus of NGC~2617 remains in a high state and can still be classified as a type-1 AGN \citep{oknyansky9015}. The optical photometry and IR photometry show that the activity of NGC~2617 is continuing and that it underwent another series of outbursts in 2016 April--June.  These outbursts are comparable, in level, to those when NGC~2617 was observed by  \cite{shappee2014} in 2013 May \citep{oknyansky9030,oknyansky9050}. We subsequently applied for soft X-ray and UV observations with the {\it Swift}/XRT.  These began on 2016 May 17 and continued till 2016 June 23.

\subsection{Optical spectroscopy}

\begin{figure}
	\includegraphics[width=\columnwidth]{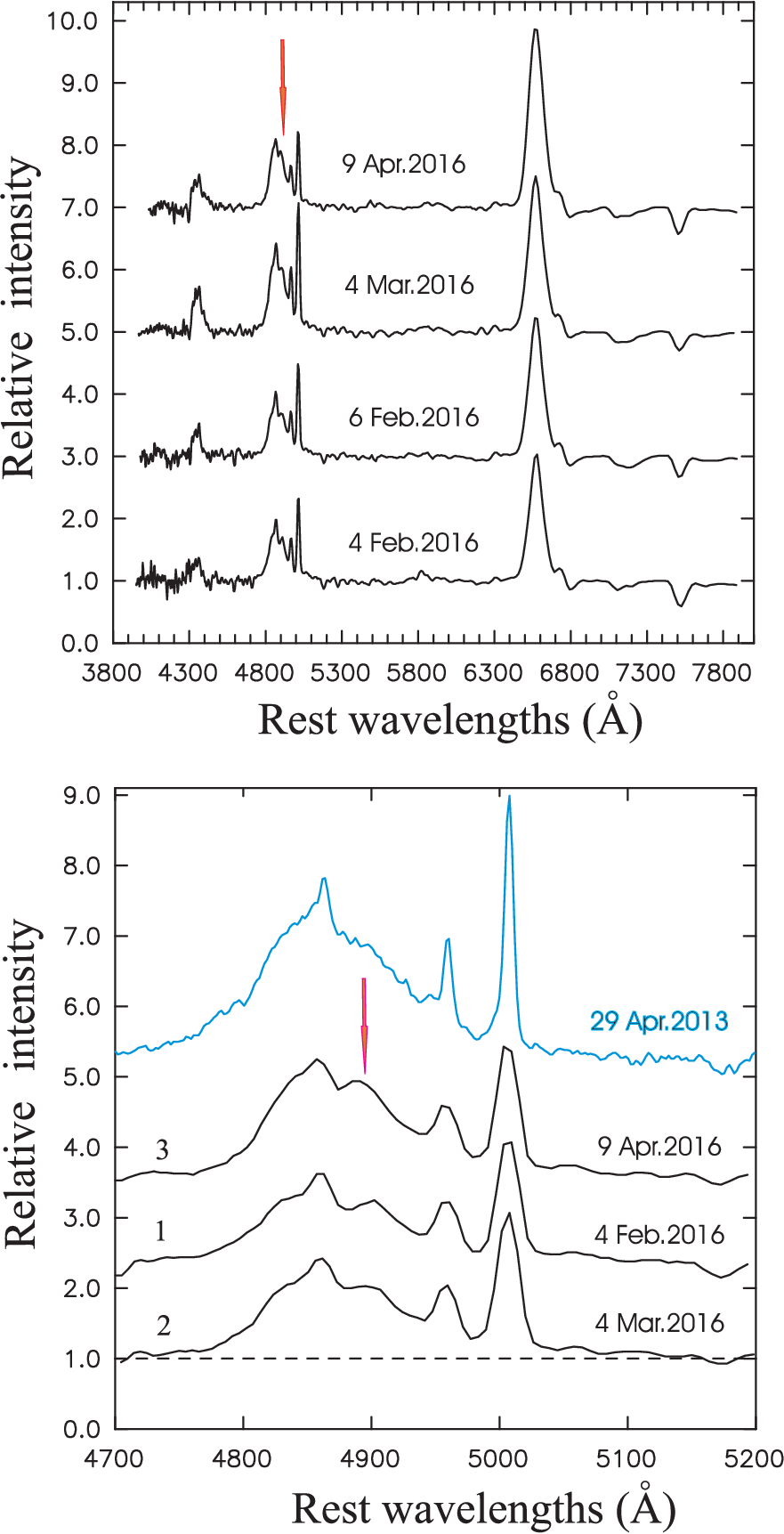}

    \caption{Top panel: our mean spectra for four dates normalized to the continuum and offset  for clarity. Bottom panel: comparison between the \citet{shappee2014} Apache Point Observatory 2013 April 25 spectrum of the H$\betaup$ region of NGC~2617 and our spectra from 2016 February 4 (1), March 4 (2) and April 9 (3). Spectra have been normalized to the continuum level, and  then relative calibration has been performed by assuming a constant flux in [\ion{O}{iii}] $\lambda$5007 and $\lambda$4959.  Spectra are offset for clarity. The arrows show the locations of the displaced emission peak in the red wing of H$\betaup$}
    \label{fig1}
\end{figure}

We obtained optical spectra covering 4100--7000 \AA\ with the 2$\times$2 prism spectrograph and a 4K CCD (spectral resolution 3--7~\AA) on the 2-m Zeiss telescope of the Shamakhy Astrophysical Observatory (ShAO) on the four nights of 2016 February~3 and 4, March~4 and April~9. Examples of mean spectra of the H$\betaup$ region for three of the nights can be seen in Fig.~\ref{fig1} together with a spectrum from \cite{shappee2014} .   It can be seen from all our 2016 spectra that NGC~2617 can be classified  without any doubt as a type-1 AGN.

In our spectra one can see a  displaced emission component in the red wing of H$\betaup$ at a relative velocity of $\sim$ +2500 km s$^{-1}$ which was not apparent in spectra obtained in 2013.  We could not verify that this new component is also present in  the H$\alphaup$ profile because of the inferior resolution of the prism spectrograph at long wavelengths. The emission component cannot be identified confidently in the profile of H$\gamma$ as there is strong [\ion{O}{iii}] $\lambda$4363 emission.

\subsection{I\uppercase{R} photometry}

\begin{figure*}
	\includegraphics[scale=0.8,angle=0,trim=0 0 0 0]{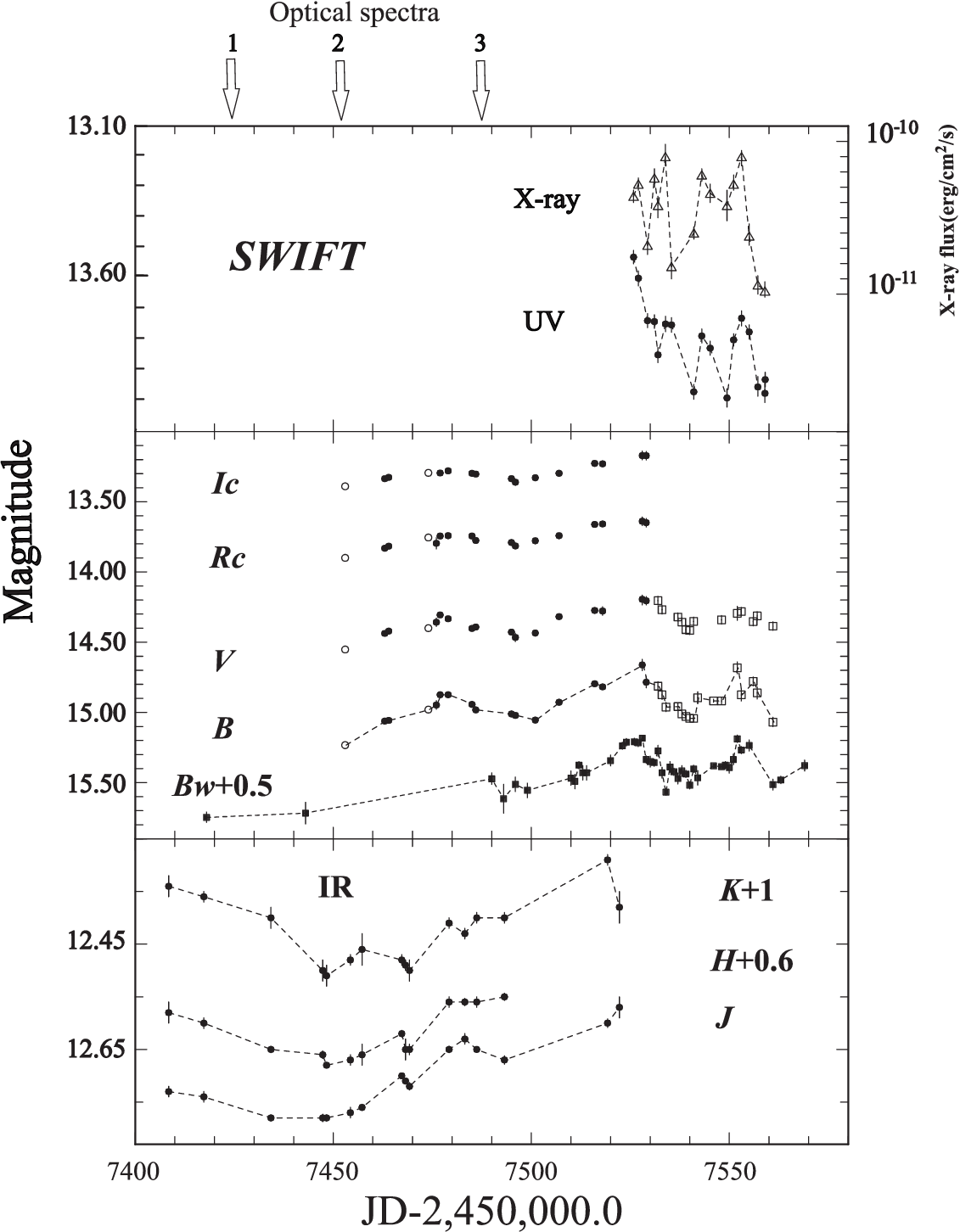}
    \caption{Near--IR (bottom panel), optical (middle panel) and UV--X-ray (top) photometric observations of NGC~2617 over the $\sim$ 5-month period from 2016 January 30 to June 29.  The solid circles in the middle panel are $BVR_cI_c$ data obtained with AZT-5 while the open circles are observations with the Zeiss-600.   Open boxes are filtered $BV$ data obtained by the MASTER network, and solid squares ($Bw$) are from unfiltered MASTER data reduced to the $B$ system.
   In the top panel the solid circles are the combined {\it Swift} UV photometry reduced to the $UVW1$ system whilst the X-ray flux is shown by the open triangles.
   Error bars are shown, but they are generally smaller than the plotting symbols. The dates of the optical spectra are indicated.}
    \label{fig2}
\end{figure*}

The $JHK$ photometry was obtained from 2016 January 20 to May 13 with the new 2.5-m telescope of the SAI Caucasus Mountain Observatory equipped with the ASTRONIRCAM instrument (an MKO photometric system with the HAWAII 2-RG detector) operating in the $JH$ and $K$ bands.  The data were calibrated using 2MASS stars within 1.5 arcmin of NGC~2617 and then converted into the MKO system following \citet{leggett2006}.  We  performed photometry using 2.5 and 5 arcsec radius apertures and found that the signal (ratio of real variations to noise) was better for the smaller aperture.  Measurements with a 5 arcsec radius aperture are useful, however, for comparison with the IR photometry of \citet{shappee2014} who use the same size photometric aperture. We measured the background with an annulus of radii 30--60 arcsec.  The $JHK$ light curves for the 2.5 arcsec radius aperture can be seen in Fig.~\ref{fig2}.  As we discuss below, there are some difference in the variations of different IR bands because of time lags.

\subsection{Optical $BVR_cI_c$ observations}

We obtained optical $BVR_cI_c$ CCD data with the AZT-5 (a Maksutov 50-cm meniscus telescope equipped with an Apogee Alta U8300 CCD camera) at the MSU Crimean Observatory and with the Zeiss-600 using a 4K CCD at ShAO on 17 nights in 2016 March--May.  The data were calibrated using SDSS stars within 1.5 arcmin of NGC~2617 and transformed into the Johnson--Cousins magnitude system in the same manner as \citet{shappee2014}.  We measured the background within an annulus of radii 35--45 arcsec.

The $BVR_cI_c$ light curves are shown in Fig.~\ref{fig2}.  The magnitudes are for an aperture of 5 arcseconds radius. NGC~2617 can be seen to have  brightened by about 0.3 mag in  $B$ during March and then decreased by 0.1 mag in the middle of April.   At the end of April it began to brighten again and reached a maximum of $\sim 14.6$ in $B$ on May 19.   One more maximum of the similar magnitude  was reached near June 15.    The amplitude of variations in $V$ was about half the amplitude in $B$. Variations in  $R_c$ and $I_c$ were synchronized with $B$ band variations but still with smaller amplitudes.

\subsection{Observations with the \uppercase{MASTER} network}

\begin{figure*}
\includegraphics[scale=1.0,angle=0,trim=0 0 0 0]{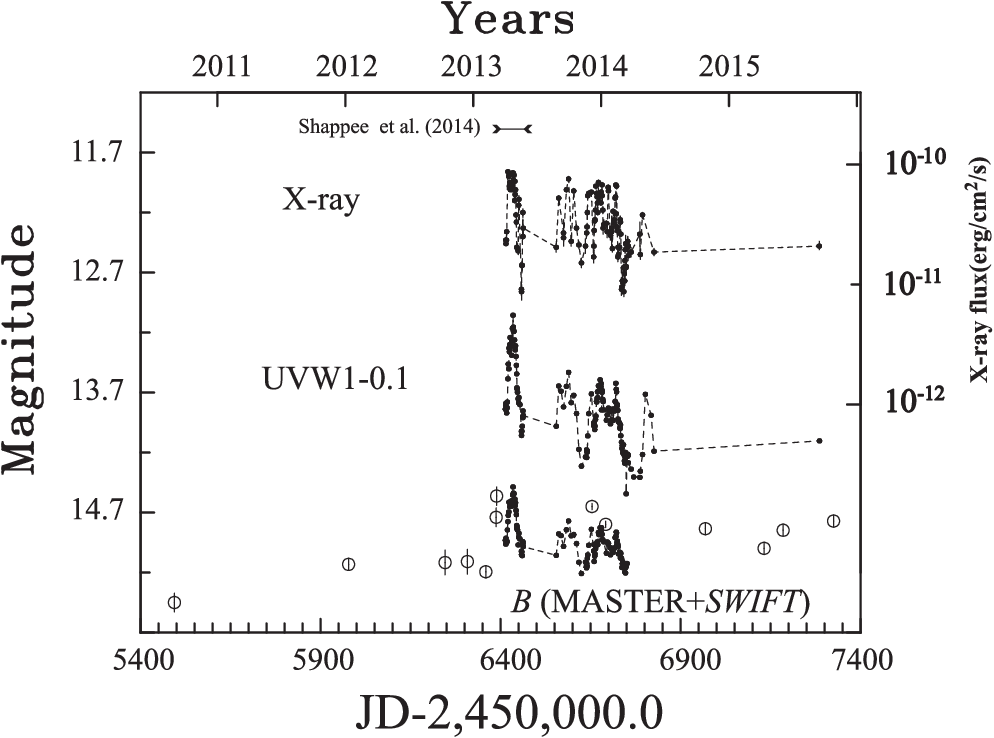}
 \caption{Bottom panel: unfiltered MASTER optical photometry of NGC~2617 reduced to the $B$ system (large open circles) and $B$ data obtained by {\it Swift} (closed circles).  Middle panel: UV photometric observations in $UVW1$.  Top panel: X-ray flux obtained by {\it Swift} (filled circles). The period of observations of \citet{shappee2014} is indicated.}
    \label{fig3}
\end{figure*}

We also give an optical light curve for 2010--2016 from the MASTER Global Robotic Network \citep{lipunov2010}.  The white light (unfiltered) magnitudes,  $\Delta{W}$, relative to two nearby comparison stars, correspond to a photometric aperture  of 7.5 arcsec (4 pixels) radius.   The variations are similar to those shown by \citet{shappee2014} for 2012--2013.  NGC~2617 had about the same brightness in 2012 and in 2016 February,  and in 2016, it has a type 1 spectrum, so we suggest that the dramatic change in type of NGC~2617 was not  related to the brightening observed in 2013 but probably happened between 2010 October and 2012 February.  From the end of April we obtained observations of NGC~2617 as often as possible. The light curve for the unfiltered photometry correlates well with our $BV$ photometry (see Fig.~\ref{fig2}). From  2016 May 16 we started observations in the $B$ and $V$ bands, mostly using the MASTER telescopes located in South Africa (SAAS) in order to be able to continue with our $BV$ light curves. These measurements were made with the same aperture parameters as for our previous $BVR_cI_c$ photometry. We have combined these estimates in Fig.~\ref{fig2}.  Since variations in $BVR_cI_c$  and the unfiltered MASTER photometry are well correlated, we reduced $W$ magnitudes to the $B$ system and called them $B_{w}$ (see Fig.~\ref{fig2} and~\ref{fig3}).

\subsection{{\it Swift} ToO optical, ultraviolet, and X-ray observations}

The {\it Swift} observatory monitored NGC~2617 more or less regularly for a year from 2013 May. Some of these observations (those of 2013 April--June) were published by \cite{shappee2014}. In addition to using the available data in the {\it Swift} archive we applied for a ToO programme for 2016 May 17 through June 20 2016. All data were reduced uniformly to trace the evolution of the behaviour of NGC~2617 on a longer time--scale.

The XRT telescope observed NGC~2617 both in photon counting  and in windowed timing  modes depending on its brightness. The spectra were reduced using the online tools provided by the UK {\it Swift} Science Data Centre \citep[\url{http://www.swift.ac.uk/user_objects/};][]{2009MNRAS.397.1177E}. The spectra obtained were grouped to have at least one count per bin
and were fitted in the 0.3--10 keV band with the {\sc XSPEC} package using Cash statistics \citep{1979ApJ...228..939C}.  Following \cite{shappee2014}
we used a simple absorbed power-law model for the fitting procedure with the absorption parameter frozen at the Galactic value ($N_{\rm H}=3.64\times10^{20}$ cm$^{-2}$; \citealt{2005A&A...440..775K}).

The {\it Swift} Ultraviolet/Optical Telescope  observes in parallel with the XRT telescope thus making it possible to get a
broad--band view from the optical to X-rays. During our programme the observations were made with the `filter-of-the-day'. Analysis of images
has been done following the procedure described on  the web-page of UK Swift Science Data Centre. Photometry was performed with the {\tt
uvotsource} procedure with source apertures of radii 5 and 10 arcsec for the background (with the centre about 1 arcmin
away from the galaxy) for all filters.

Given the very high correlation between variations in different UV bands in previous data we reduced the new UV data to one band, $UVW1$, and we show the combined UV light curve together with new X-ray flux variations in Fig.~\ref{fig2}.  The old X-ray flux, $UVW1$ and $B$  variations in combination with unfiltered MASTER optical data (2010--2015) are shown in Fig.~\ref{fig3}. As can be seen in the figure, all variations in the optical, UV and X-ray are correlated. The amplitude is largest for the X-ray and decreases with increasing wavelength. Possible lags between these variations are discussed below.

\section{Time lags}
We measured lags between continuum variations in different bands
using two independent methods: MCCF, a modification of traditional cross-correlation techniques (see \citealt{oknyanskii1993}) and a Bayesian analysis using the JAVELIN software 

\subsection {MCCF}
Cross-correlation analysis of astronomical time series is complicated because the sampling is usually uneven. As a rule, series of astronomical observations inevitably have gaps because of seasons when the object is invisible and interruptions due to the full moon, weather, and observing schedules etc. Classical cross-correlation analysis methods were developed only for uniformly sampled time series.  The analysis of non-uniform astronomical series requires special techniques. In practice, a number of methods can be simply divided into two types:
(1)	interpolation or approximation of the series in the time domain, followed by the application of classical methods of cross-correlation analysis. This is the approach  of the widely-used interpolation method ICCF ( \citealt{Gaskell86});  and (2)  methods that do not interpolate in the time domain but bin the correlation function by lag.  A typical example of this type of methods is the discrete correlation function \citep[DCF;][]{Edelson88}.

In order to improve the methodology of ICCF and DCF, we used an intermediate method, which is between the  ICCF and DCF methods, and we call it MCCF (\citealt{oknyanskii1993}).  At the heart of our method is the ICCF method, but we strive to reduce the contribution of the error interpolation by introducing a maximum interval, MAX, used for interpolating points. We use only those interpolated points that are separated in time from the nearest observation points by no more than the value of MAX. Accordingly, at each lag we will typically have a different number of pairs of points to calculate the correlation while in the ICCF method  this number remains fixed.  The value of MAX is chosen to be as small as possible and yet make it possible to calculate the cross-correlation function in the range of lags we are interested in (for more details see \citealt{oknyansky2014}).

To estimate the errors in the derived lags, we applied
the same Monte Carlo procedure as before (\citealt{oknyanskij1999}; \citealt{oknyansky2014}). We simulated artificial series (10000 times) with the same temporal distribution, power spectrum
and measurement errors as the observed data. At the same time we introduced the same time lag between the time series as found from the real observations.
We then plotted the distribution of lags. In addition,
as before, as a check, the error was estimated using
the analytical formula given in \citealt{Gaskell87} (see their equation 4). As has been noted before
(see \citealt{Koratkar91}; \citealt{Maoz91}
and \citealt{oknyanskij1999}), error estimates for the
Monte Carlo method and this analytical formula are in
agreement.  We also estimate a 95\% confidence  level  for each lag by performing 10,000 independent simulations the same way as above, but for two sets of data without any correlation and lag. We then estimate single-tail 97.5\% confidence levels for negative and positive variations of the correlation coefficients.

Using the MCCF method we get cross-correlation functions, $r(\tau)$, as a function of the lag, $\tau$, for the time series $X(t)$ and $Y(t)$.  From $r(\tau)$ the  best lag  between   $X(t)$ and $Y(t)$ can be estimated in at least two ways: from  $\tau_{peak}$  corresponded to  $r_{max}$, and from the centroid, defined as $\tau_{\rm cent} = \int
\tau r(\tau)\,d\tau / \int r(\tau)\,d\tau$ for values of $r \geq 0.8 r_{\rm
  max}$.

\subsection {JAVELIN}

We  also used the reverberation mapping package  JAVELIN  \citep[\url{http://bitbucket.org/nye17/javelin};][]{zu2011,zu2013};  formerly known as SPEAR). \citet{Gaskell87} showed that AGN light curves can be represented by a damped random walk. JAVELIN models the continuum light curve as a damped random walk, assumes a transfer function, and then fits
this $~10^4$ times to both time series. In place of the usual cross-correlation function JAVELIN produces a histogram of lags. These histograms can be used for the estimation of optimal mean  lags $\tau_{JAVALIN}$ and its $1 \sigma$ intervals (see thre details in e.g., \citealt{shappee2014} ; \citealt{zu2013}).

\subsection {Results}

\begin{figure}
\includegraphics[width=\columnwidth]{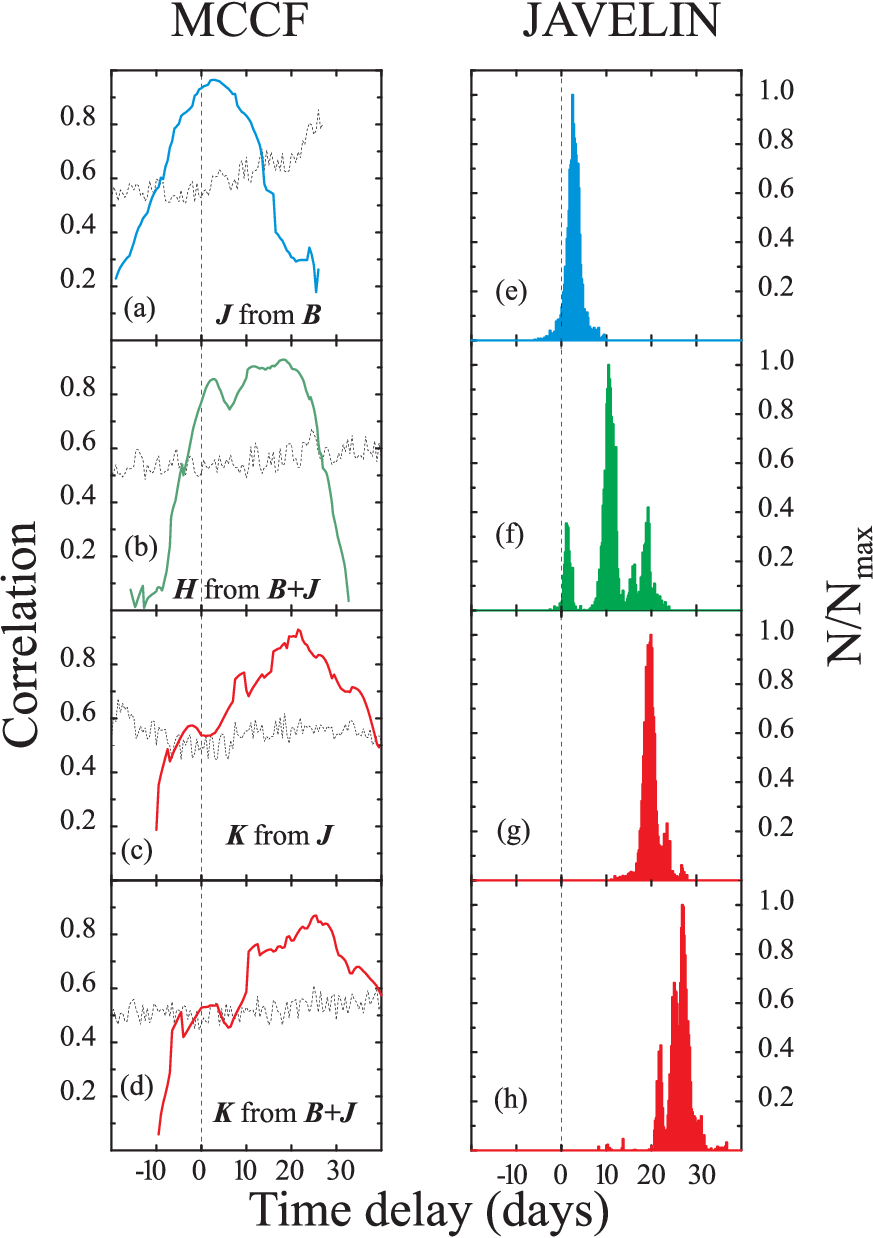}
    \caption{Left--hand  panels: CCFs calculated by the MCCF method of \citet{oknyanskii1993} for our 2016 IR ($JHK$) and optical ($B$ band) photometry The dashed  lines are 97.5\% single-tail confidence levels estimated via  Monte Carlo simulations (see the text).  Right--hand panels: the corresponding lags obtained from the JAVELIN method of \citet{zu2011}. The histograms are normalized by dividing by their maximum values. The vertical dashed lines indicate zero lag.
}
    \label{fig4}
\end{figure}

\begin{figure}
	\includegraphics[width=\columnwidth]{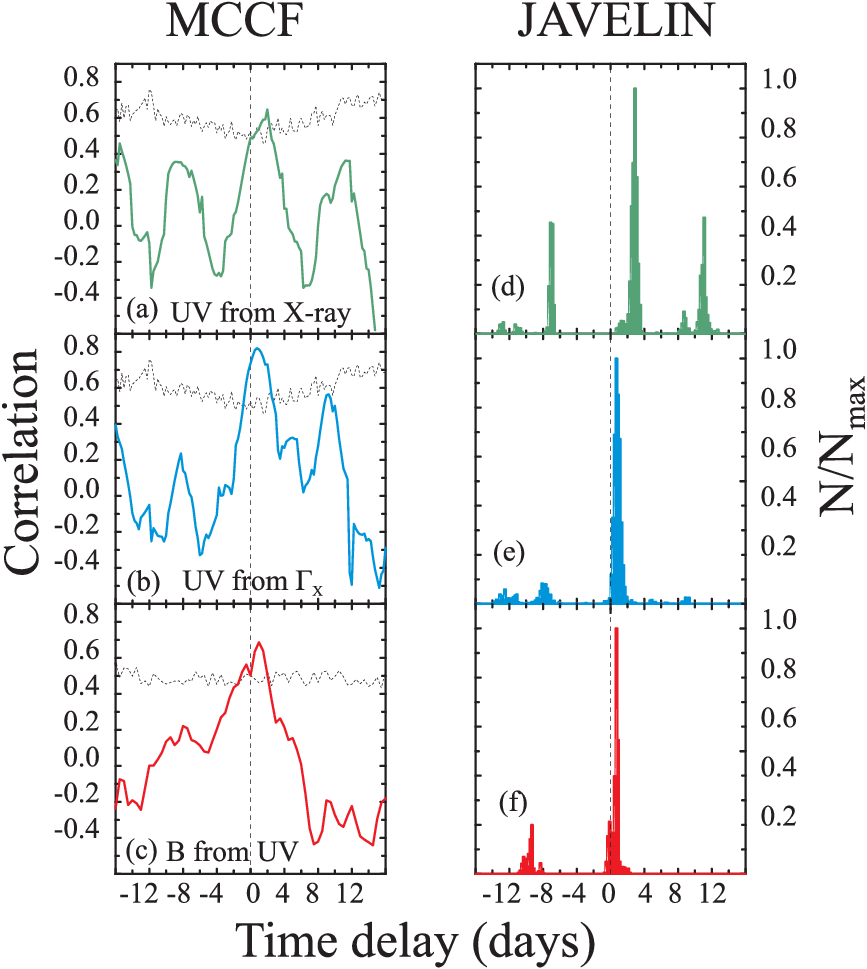}
    \caption{Left panels: CCFs for our May and June 2016 {\it Swift} data.  CCFs have been calculated by the MCCF method. Panel (a) shows the MCCFs for $UVW1$ lagging the X-ray flux, (b) shows $UVW1$ lagging the X-ray photon index, $\Gamma_\mathrm{X}$, and (c) shows the unfiltered MASTER data lagging $UVW1$. The dashed lines are 97.5\% single-tailed  confidence levels obtained using Monte-Carlo simulations.  Right panels: the corresponding lags obtained using JAVELIN. The histograms are normalized by dividing by their maximum values. The vertical dashed lines indicate zero lag.
}

    \label{fig5}
\end{figure}

\begin{figure}
	\includegraphics[width=\columnwidth]{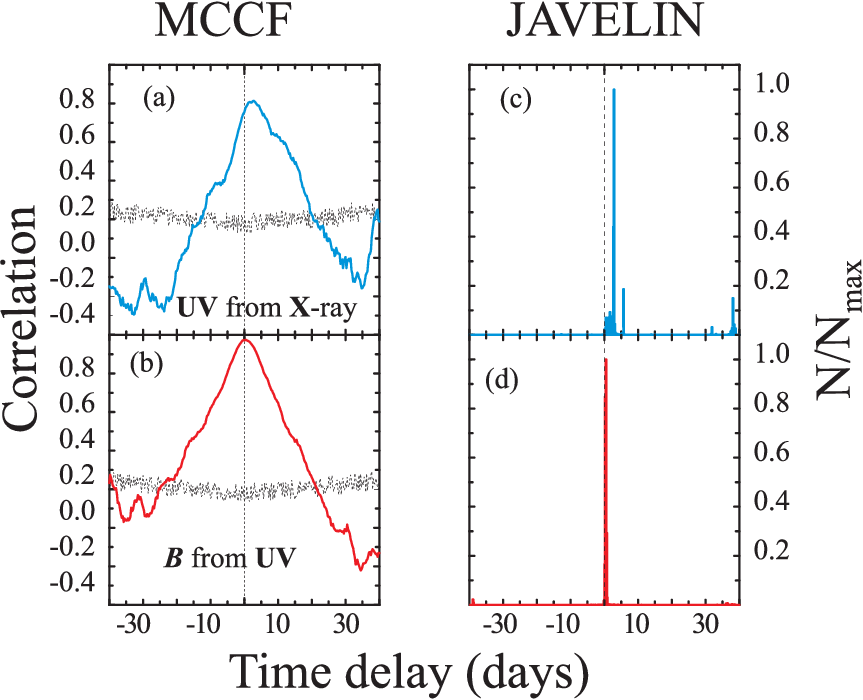}
    \caption{Left--hand panels: CCFs for archival {\it Swift} data from 2013 to 2014.  (a) shows the MCCFs for $UVW1$ lagging the X-ray flux and(b) shows the CCF for the $B$--band lagging $UVW1$. The dashed lines are 97.5\% single-tail confidence levels obtained from Monte Carlo simulations. Right--hand panels: corresponding lags obtained from JAVELIN. The histograms are normalized by dividing by their maximum values. The vertical dashed lines indicate zero lag.
}
    \label{fig6}
\end{figure}

Results of our reverberation mapping analyses  for IR, optical, UV and X-Ray data using  MCCF and JAVELIN are presented in Figs~\ref{fig4}--\ref{fig6} and Table~\ref{tab1}.  A quick inspection of Figs~\ref{fig4}--\ref{fig6} confirms that  the results obtained with  these two  different   methods are mostly consistent with each other. Some differences in the lags obtained with the two different methods are, in  most of the cases, in the limit of estimated errors (see e.g., \citealt{Lira2015}).  The difference and advantages of  peak and centroid lags have been  discussed many times in the literature (see e.g. \citealt{Koratkar91}; \citealt{Peterson2004}). An advantage of the centroid is that it is better defined when the cross-correlation function (CCF) has a broad peak. As can be seen from Figs~\ref{fig4}--\ref{fig6}, in most of the cases we have principal peaks in the cross--correlation functions that  are isolated and well defined. In these cases  we prefer the peak lags that are less dependent on free parameters. In what follows we  will talk mostly about the peak lags, but the centroids from MCCF and the JAVELIN values for lags can be seen in Table~\ref{tab1}.

Our analysis shows [see Figs~\ref{fig4}~(a) and (e) and Table~\ref{tab1}] that variations in the $J$ band lag those of the $B$ band by about 3 d. We take this lag into account  and use a linear regression relationship to reduce the $J$ fluxes to the $B$ system and then combine the reduced data with $B$. This combined data set  we will refer to as the `$B+J$'~data.   The variations of the $H$ band lag those of $B+J$ by about 18 d. [see Figs~\ref{fig4}~(b) and(f) and Table~\ref{tab1}]. The $K$ band  lags the $B$ band by about 21.5 d. and lag the $B+J$ by about 25.3 d. (see Figs.~\ref{fig4}~(c),(d),(g) and (h)).  We interpret the near-infrared (NIR) lag of the $K$ band as being due to thermal reradiation of shorter wavelength flux by  hot dust.  We suggest that the shorter lag of $\sim$ 3 d. for the $J$ band, compared with the $B$ band, is probably because of the strong contribution of emission from  the outermost part of the accretion disc. The lag for $K$ behind $B$ is therefore about 25$\pm$3 d.   This is more than twice the lag found by \citet{shappee2014} of about 9 d, but an examination of  the IR data shows that the light curve for $K$  in 2013 is just line trend at a confidence level of 0.99 and is not appropriate for reverberation mapping analysis. The reason why lags of about 8--10 d. were found by Shappee et al. is  related to the properties  of the optical  and UV light curves in 2013. They have an approximately linear  rise and  then a roughly linear  decrease. The maximum of these light curves is shifted relative to the end  of the IR data by about 8--10 d. So, for lags more than  8--10 d. the correlation is very high.  For smaller lags the correlation will be lower.  Shappee et al.\@ forced JAVELIN to return only lags of less than 10 d. Given the limitations of the data we can only  say that the the lag for $K$ from UV and $B$ variations is not shorter than about 8~d. So our analysis does not confirm the result of Shappee et al.\@ that  the lag for $K$ relative to the UV is about 9 d. It is quite likely that the lag changed during the three previous years, but available data  cannot establish this yet.

Our cross-correlation analyses for optical, UV and X-ray data are presented in  Fig.~\ref{fig5} for the new data for 2016 and in Fig.~\ref{fig6} for the data from 2013 to 2014.  We find that in 2016 (see Fig.~\ref{fig5}) optical variations probably lag the UV by about one day or less. The UV variations themselves have a lag about 2--3 d. behind the X-ray flux and about 1 d. behind the photon index, $\Gamma_\mathrm{X}$. For the older 2013--2014 data, UV variations in different bands correlate very well with each other (correlation coefficients are about 0.98 and the lags are consistent with zero) and have very similar MCCFs with X-ray variability showing a lag of about 2.5 $\pm$ 0.5 d. In Figs~\ref{fig6} (a) and (c) we therefore show results only for $UVW1$.   Variability in the $B$ band correlates well with UV variability with lags of about 0.5 d.
	
The lags we obtain are consistent with the light-travel times expected when UV and optical radiation are driven by the reprocessing of extreme UV and soft X-ray emission. If some of the changes in lags between 2013--2014 and 2016 are real, they could be  due to changing physical conditions in the object.

\begin{table}
\centering
\caption{Lags from the reverberation mapping analysis with the MCCF and JAVELIN methods in days. MCCF and JAVELIN $1\sigma$ confidence limits are presented.}
\begin{tabular}{cccccccc} \hline

  & MCCF  &    &JAVELIN        & Time       \\ \hline \smallskip
  & $ \tau_{peak}$  &  $\tau_{cent}$ &  $\tau_{JAV}$                &\\          \\ \smallskip
$J$ from $B$ & $2.8^{+1.2}_{-0.8}$  &$3.0^{+1.0}_{-1.2}$&  $2.7^{+1.2}_{-1.4}$ & 2016 \\ \smallskip
$H$ from $J+B$ &$18.2^{+3.0}_{-4.0}$    & $13.9^{+4.0}_{-3.0}$& $10.7^{+1.2}_{-1.2}$ &     --        \\ \smallskip
$K$ from $J$ &  $21.5^{+2.4}_{-2.6}$  &$20.5^{+2.2}_{-2.4}$& $19.9^{+1.4}_{-1.6}$ &--               \\ \smallskip
$K$ from $J+B$ &  $25.3^{+3.4}_{-2.3}$&$24.5^{+3.1}_{-2.6}$& $26.0^{+2.1}_{-2.1}$ & --              \\ \smallskip
$UV$ from $X_{ray}$ & $2.0^{+0.7}_{-0.5}$    & $1.5^{+0.7}_{-0.5}$& $2.7^{+0.4}_{-0.2}$ & --               \\ \smallskip
$UV$ from  $\Gamma_\mathrm{X}$  & $0.8^{+0.5}_{-0.5}$  & $0.9^{+0.4}_{-0.2}$& $0.8^{+0.3}_{-0.3}$& --       \\ \smallskip
$B_W$ from $UV$ & $1.0^{+0.5}_{-0.5}$    &$0.5^{+0.5}_{-0.5}$& $0.6^{+0.3}_{-0.5}$ & --               \\ \smallskip
$UV$ from $X_{ray}$ & $2.5^{+0.5}_{-0.7}$     &$2.9^{+0.5}_{-0.8}$& $2.5^{+0.6}_{-0.8}$ & 2013--2014               \\ \smallskip
$B_S$ from $UV$ &$ 0.4^{+0.4}_{-0.4}$  &$0.6^{+0.3}_{-0.3}$& $0.4^{+0.2}_{-0.2}$ & --               \\

\hline

\end{tabular}

\label{tab1}
\end{table}

\section{Discussion}

NGC~2617 is one of the clearest cases where  appearances,  both as  type-1 and as type-2 AGNs have been manifested in the same object at different epochs. What must happen to make such a dramatic change possible?  CL AGNs like NGC~2617 present problems for the simplest `SPM' unification models where type--2 and type--1 AGNs are the same and the difference in type is only due to the orientation of the observer.  In this model we see a type-2 AGN if the BLR and accretion disc are blocked from our view by obscuring dust surrounding the AGN perpendicular to the axis of symmetry \citep{Keel80}.  Orientation obviously cannot change on the timescale of the observed type changes and hence some other explanation is needed.

It is difficult to fit our observations with transient events such as the tidal disruption of a star \citep{eracleous1995} -- a so-called `tidal disruption event' (TDE) -- or a supernova in the nucleus of NGC~2617 (as was  considered for the CL AGN NGC~7582 by  \cite{aretxaga1999}).  Supernovae do not repeat and produce outbursts lasting several years.  They also produce smooth, monotonically decaying light curves that are quite different from what we find for NGC~2617. The changing appearance of NGC~2617 is obviously different from NGC~7582 and is more similar to the sorts of changes see in  NGC~863 \citep{Denney2014}) and Mrk~6 \citep{pronik_chuvaev1972,chuvaev1991}, which spent a long time in type-1 states  after transitions from  type-2 states. The main problem with invoking a TDE is that they are extremely rare.  The  presence of a secondary supermassive black hole could significantly increase the tidal disruption rate (see, e. g., \citep{Ivanov2005}). However, at present there is no evidence of a secondary black hole in NGC~2617.  Like supernovae, a TDE near a  quiescent black hole produces a smooth, decaying light curve.  The case of a TDE near a black hole in an AGN could be different because of the preexisting accretion disc.  A TDE could perhaps produce recurrent brightening for some time by perturbing the disc and adding material to it.

We consider it more likely that AGN type changes are the result of normal AGN processes.  It has also long been argued that AGN type depends on the accretion rate \citep{Dibai81} and hence  changes in type arise from variations in the luminosity of the central engine \citep{lyutyj1984}. The multiwavelength observations of NGC~2617 support this picture.  To explain changes in the optical obscuration of AGNs and related changes in X-ray spectra it has been proposed that the obscuring material has a patchy distribution (see \citealt{Turner+Miller09} for a review).

We propose that AGN type changes arise from a {\em combination} of a change in central luminosity {\em and} a related change in the obscuration.  \citet{Heard+Gaskell16} show that the substantial extinction seen in a large fraction of AGNs mostly comes from dust  between the narrow-line region and the BLR.  The surprising wavelength independence of IR lags in reverberation mapping studies of many AGNs \citep{oknyansky2015} can be explained by the hot dust being in a hollow, biconical outflow.  When the level of activity increases, the extra radiation will sublimate the nearest dust in the biconical outflow.  Such changes in lags in the  NIR have been seen in several AGNs (see \citealt{oknyansky2015}).  As a result of the destruction of inner dust, we can get a less obstructed view of the BLR and the AGN will be seen as a type 1.  Sooner or later the level of activity will decrease again, the dust will re-form, and the AGN will again be seen a type 2.

 \cite{Barvainis1992} has considered  `reformation'  and `survival'  models where dust grains are located in the clouds. These models can be reconciled with
observations by introducing an axially symmetric distribution of these clouds
with a strong concentration to the equatorial plane (see, e.g., \citealt{Rowan-Robinson1995}).  For this type model we can use improvements from \cite{Sitko1993}:
\begin{equation}
r_{evap}=136~L_{UV,44}^{0.5}~[T/1500K]^{-2.8}~[0.05/A_{\mu}]^{0.5}~ e^{-\tau/2}~ $light-days$
\end{equation}
where $r_{evap}$ is the sublimation radius,  $A_{\mu}$ is the graphite grain size in $\mu$, and $\tau$ is the optical depth of
the clouds to UV radiation.  Following Sitko et al., \cite{oknyanskij2001} use the same parameters: $T=1700$ K, $A_{\mu}=0.15$, $\tau=1$  and find a good agreement with observed IR lags for 11 AGNs. If  for NGC~2617 we take $L_{uv} = 7\times 10^{43}$ erg/s  ( $\sim$ 10\% of $L_{\rm{Edd}}$ for $M_{\rm BH} $=$ 4 \times 10^7 {\rm{M}_\odot}$  \citealt{shappee2014}) we get a  rough estimate for the sublimation radius at NGC2617 of about 32 light-days.
NIR reverberation radii have been shown to scale with the central engine's
luminosity $L$ as $\sim L^{1/2}$ (\citealt{oknyanskij2001}; \citealt{Suganuma2006}), which suggests that the properties of the innermost dust grains are indeed similar in different objects
(\citealt{Barvainis1992}).  The radius does not necessarily show the expected  $L^{1/2}$  dependence for an individual object. For example, for NGC~4151  reverberation  mapping shows that the changes in radii have a lag of several years  relative  to changes in $L$ (\citealt{oknyanskij2008};\citealt{Kishimoto2013}).  So, if the luminosity in NGC~2617 increased sharply in 2013 the expected  change in the NIR reverberation radius could be observed a few years later.

The appearance and disappearance of emission components in the profiles of broad Balmer lines, such as we report here for NGC~2617, are common in type-1 AGNs including CL AGNs (e.g., NGC~5548 \citealt{Bon+15};  NGC~4151 \citealt{oknyanskij1999b}).  There is no consensus yet on the cause of these features. Possibilities considered include binary supermassive black holes (see \citealt{Gaskell10}), elliptical discs (which could arise when a star is disrupted, \citealt{eracleous1995}), regions of enhanced emission in a disc \citep{zheng1991}, and off-axis illumination \citep{Gaskell11,Goosmann+14}.

Previous studies \citep{Breedt2008,shappee2014,Connolly2015,Noda2016,Buisson2017} confirm that strong X-ray UV/optical correlations with short (no more than few days) lags for longer wavelengths  are the norm in radio-quiet AGNs rather than the exception, which seems  to support  suspicions that the UV/optical variations are primarily produced by the reprocessing of variable EUV/soft X-ray emission by the accretion disc. However, in several cases correlations  are present on short time--scales but absent on long time--scales. For some objects, it is known that  the level  and type of correlation  can be changed  significantly  from one epoch to another (for example,  NGC3516 \citealt{Noda2016}). For NGC2617 we find a strong X-ray--UV/optical correlation during 2013--2014  but the long-term variations for these wavebands are not exactly the same as what can be seen in  Fig.~\ref{fig3}. For 2016 May--June  the correlation is not as strong  as it was for 2013--2014. This could be  caused by the existence of some uncorrelated components in the variations.  Following Breedt et al. (2008), we argue that the data require a combination of X-ray reprocessing and independent UV/optical and X-ray long-term components probably connected to accretion rate variations in the accretion disc or some other mechanism (see the discussion in \citealt{Breedt2009}).  Additionally, the existence of uncorrelated components in the X-ray and UV/optical variations could be explained by some variable obscuration on the line of sight with differences in the locations  and sizes of the regions. Some previous studies (e.g. \citealt{shappee2014}; \citealt{Noda2016})  have noted that the lags found from correlations of the X-ray--UV/optical variations are longer by a factor of  a few than can be expected from the standard thin disc model.  \citet{Gaskell15} has proposed that this is because the luminosity has been underestimated because of the neglect of reddening.  Other possible explanations have been proposed to explain the discrepancy (see, e.g., \citealt{Buisson2017}) including  the effect of emission lines and the Balmer continuum  (\citealt{Korista2001}).

Observations presented here mostly support a scenario where the variability across several wavebands (spanning X-rays--NIR) is   driven by variable illumination of the  accretion disc by soft X-rays (Shappee et al. 2014). However, we prefer to interpret the IR radiation  (near 2.2 $\mu$)  as being dominated by thermal re-radiation of the dust in clouds located farther away from the black hole than the AD  and the outer edge of the  BLR.

\section{Summary}

We have shown using spectroscopy and multi-wavelength photometry that NGC~2617 continues to be in a high state and to appear to be a type-1 AGN.  The duration of the high state and the continuing variability are not readily consistent with the type change being due to a tidal disruption event or a supernova.  We propose that the change of type is a result of increased luminosity causing the sublimation of dust in the inner part of the biconical dusty outflow.  This leads to a much more direct view of the central regions.

\section*{Acknowledgements}
We are grateful to B.~Shappee for sending us data in a digital format. We thank the SAI director, A.~Cherepashchuk, and ShAO director, N.~Jalilov, for granting us Director's Discretionary Time for observations, and the staff of the observatories.  We also express our thanks to N.~Gehrels for approving the {\it Swift} ToO observation requests and the {\it Swift} ToO team for promptly scheduling and executing our observations. We are grateful to P.~B.~Ivanov for useful discussions  and to the referee for useful comments that have improved the presentation of this paper.  This work was supported in part by the M.~V. Lomonosov Moscow State University Program of Development, in part by the National Research Foundation of South Africa, in part by the Russian Foundation for Basic Research through grant 14-02-01274 and  in part by Russian Science Foundation through grant 16-12-00085.




\bibliographystyle{mnras}
\expandafter\ifx\csname natexlab\endcsname\relax\def\natexlab#1{#1}\fi
\bibliography{ngc2617} 

\begin{thebibliography}{}
\makeatletter
\relax
\def\mn@urlcharsother{\let\do\@makeother \do\$\do\&\do\#\do\^\do\_\do\%\do\~}
\def\mn@doi{\begingroup\mn@urlcharsother \@ifnextchar [ {\mn@doi@}
  {\mn@doi@[]}}
\def\mn@doi@[#1]#2{\def\@tempa{#1}\ifx\@tempa\@empty \href
  {http://dx.doi.org/#2} {doi:#2}\else \href {http://dx.doi.org/#2} {#1}\fi
  \endgroup}
\def\mn@eprint#1#2{\mn@eprint@#1:#2::\@nil}
\def\mn@eprint@arXiv#1{\href {http://arxiv.org/abs/#1} {{\tt arXiv:#1}}}
\def\mn@eprint@dblp#1{\href {http://dblp.uni-trier.de/rec/bibtex/#1.xml}
  {dblp:#1}}
\def\mn@eprint@#1:#2:#3:#4\@nil{\def\@tempa {#1}\def\@tempb {#2}\def\@tempc
  {#3}\ifx \@tempc \@empty \let \@tempc \@tempb \let \@tempb \@tempa \fi \ifx
  \@tempb \@empty \def\@tempb {arXiv}\fi \@ifundefined
  {mn@eprint@\@tempb}{\@tempb:\@tempc}{\expandafter \expandafter \csname
  mn@eprint@\@tempb\endcsname \expandafter{\@tempc}}}

\bibitem[\protect\citeauthoryear{{Antonucci}}{{Antonucci}}{1993}]{Antonucci93}
{Antonucci} R.,  1993, \mn@doi [\araa] {10.1146/annurev.aa.31.090193.002353},
  \href {http://adsabs.harvard.edu/abs/1993ARA%26A..31..473A} {31, 473}

\bibitem[\protect\citeauthoryear{{Antonucci}}{{Antonucci}}{2012}]{Antonucci12}
{Antonucci} R.,  2012, Astron. Astrophy. Trans., \href
  {http://adsabs.harvard.edu/abs/2012A%26AT...27..557A} {27, 557}

\bibitem[\protect\citeauthoryear{{Aretxaga}, {Joguet}, {Kunth}, {Melnick}  \&
  {Terlevich}}{{Aretxaga} et~al.}{1999}]{aretxaga1999}
{Aretxaga} I.,  {Joguet} B.,  {Kunth} D.,  {Melnick} J.,   {Terlevich} R.~J.,
  1999, \mn@doi [\apjl] {10.1086/312114}, \href
  {http://adsabs.harvard.edu/abs/1999ApJ...519L.123A} {519, L123}

\bibitem[\protect\citeauthoryear{{Barvainis}}{{Barvainis}}{1992}]{Barvainis1992}
{Barvainis} R.,  1992, \mn@doi [\apj] {10.1086/172012}, \href
  {http://adsabs.harvard.edu/abs/1992ApJ...400..502B} {400, 502}

\bibitem[\protect\citeauthoryear{{Bon} et~al.,}{{Bon} et~al.}{2016}]{Bon+15}
{Bon} E.,  et~al., 2016, \mn@doi [\apjs] {10.3847/0067-0049/225/2/29}, \href
  {http://adsabs.harvard.edu/abs/2016ApJS..225...29B} {225, 29}

\bibitem[\protect\citeauthoryear{{Breedt}, {Uttley}, {Arevalo}, {McHardy},
  {Kaspi}  \& {Lira}}{{Breedt} et~al.}{2008}]{Breedt2008}
{Breedt} E.,  {Uttley} P.,  {Arevalo} P.,  {McHardy} I.,  {Kaspi} S.,   {Lira}
  P.,  2008, in 37th COSPAR Scientific Assembly. p.~381

\bibitem[\protect\citeauthoryear{{Breedt} et~al.,}{{Breedt}
  et~al.}{2009}]{Breedt2009}
{Breedt} E.,  et~al., 2009, \mn@doi [\mnras]
  {10.1111/j.1365-2966.2008.14302.x}, \href
  {http://adsabs.harvard.edu/abs/2009MNRAS.394..427B} {394, 427}

\bibitem[\protect\citeauthoryear{{Buisson}, {Lohfink}, {Alston}  \&
  {Fabian}}{{Buisson} et~al.}{2017}]{Buisson2017}
{Buisson} D.~J.~K.,  {Lohfink} A.~M.,  {Alston} W.~N.,   {Fabian} A.~C.,  2017,
  \mn@doi [\mnras] {10.1093/mnras/stw2486}, \href
  {http://adsabs.harvard.edu/abs/2017MNRAS.464.3194B} {464, 3194}

\bibitem[\protect\citeauthoryear{{Cash}}{{Cash}}{1979}]{1979ApJ...228..939C}
{Cash} W.,  1979, \mn@doi [\apj] {10.1086/156922}, \href
  {http://adsabs.harvard.edu/abs/1979ApJ...228..939C} {228, 939}

\bibitem[\protect\citeauthoryear{{Chuvaev}}{{Chuvaev}}{1991}]{chuvaev1991}
{Chuvaev} K.~K.,  1991, Izv. Krymskoj Astrofiz. Obser., \href
  {http://adsabs.harvard.edu/abs/1991IzKry..83..194C} {83, 194}

\bibitem[\protect\citeauthoryear{{Connolly} et~al.,}{{Connolly}
  et~al.}{2015}]{Connolly2015}
{Connolly} S.~D.,  et~al., 2015, preprint, \href
  {http://adsabs.harvard.edu/abs/2015arXiv150207502C} {} (\mn@eprint {arXiv}
  {1502.07502})

\bibitem[\protect\citeauthoryear{{Denney} et~al.,}{{Denney}
  et~al.}{2014}]{Denney2014}
{Denney} K.~D.,  et~al., 2014, \mn@doi [\apj] {10.1088/0004-637X/796/2/134},
  \href {http://adsabs.harvard.edu/abs/2014ApJ...796..134D} {796, 134}

\bibitem[\protect\citeauthoryear{{Dibai}}{{Dibai}}{1981}]{Dibai81}
{Dibai} E.~A.,  1981, Sov. Astron. Lett., \href
  {http://adsabs.harvard.edu/abs/1981SvAL....7..248D} {7, 248}

\bibitem[\protect\citeauthoryear{{Edelson} \& {Krolik}}{{Edelson} \&
  {Krolik}}{1988}]{Edelson88}
{Edelson} R.~A.,  {Krolik} J.~H.,  1988, \mn@doi [\apj] {10.1086/166773}, \href
  {http://adsabs.harvard.edu/abs/1988ApJ...333..646E} {333, 646}

\bibitem[\protect\citeauthoryear{{Eracleous}, {Livio}, {Halpern}  \&
  {Storchi-Bergmann}}{{Eracleous} et~al.}{1995}]{eracleous1995}
{Eracleous} M.,  {Livio} M.,  {Halpern} J.~P.,   {Storchi-Bergmann} T.,  1995,
  \mn@doi [\apj] {10.1086/175104}, \href
  {http://adsabs.harvard.edu/abs/1995ApJ...438..610E} {438, 610}

\bibitem[\protect\citeauthoryear{{Evans} et~al.,}{{Evans}
  et~al.}{2009}]{2009MNRAS.397.1177E}
{Evans} P.~A.,  et~al., 2009, \mn@doi [\mnras]
  {10.1111/j.1365-2966.2009.14913.x}, \href
  {http://adsabs.harvard.edu/abs/2009MNRAS.397.1177E} {397, 1177}

\bibitem[\protect\citeauthoryear{{Gaskell}}{{Gaskell}}{2010}]{Gaskell10}
{Gaskell} C.~M.,  2010, \mn@doi [\nat] {10.1038/nature08665}, \href
  {http://adsabs.harvard.edu/abs/2010Natur.463E...1G} {463, E1}

\bibitem[\protect\citeauthoryear{{Gaskell}}{{Gaskell}}{2011}]{Gaskell11}
{Gaskell} C.~M.,  2011, Balt. Astron., \href
  {http://adsabs.harvard.edu/abs/2011BaltA..20..392G} {20, 392}

\bibitem[\protect\citeauthoryear{{Gaskell}}{{Gaskell}}{2017}]{Gaskell15}
{Gaskell} C.~M.,  2017, \mn@doi [\mnras] {10.1093/mnras/stx094}, \href
  {http://adsabs.harvard.edu/abs/2017MNRAS.tmp..160G} {}

\bibitem[\protect\citeauthoryear{{Gaskell} \& {Peterson}}{{Gaskell} \&
  {Peterson}}{1987}]{Gaskell87}
{Gaskell} C.~M.,  {Peterson} B.~M.,  1987, \mn@doi [\apjs] {10.1086/191216},
  \href {http://adsabs.harvard.edu/abs/1987ApJS...65....1G} {65, 1}

\bibitem[\protect\citeauthoryear{{Gaskell} \& {Sparke}}{{Gaskell} \&
  {Sparke}}{1986}]{Gaskell86}
{Gaskell} C.~M.,  {Sparke} L.~S.,  1986, \mn@doi [\apj] {10.1086/164238}, \href
  {http://adsabs.harvard.edu/abs/1986ApJ...305..175G} {305, 175}

\bibitem[\protect\citeauthoryear{{Goosmann}, {Gaskell}  \& {Marin}}{{Goosmann}
  et~al.}{2014}]{Goosmann+14}
{Goosmann} R.~W.,  {Gaskell} C.~M.,   {Marin} F.,  2014, \mn@doi [Adv. Space
  Res.] {10.1016/j.asr.2013.11.020}, \href
  {http://adsabs.harvard.edu/abs/2014AdSpR..54.1341G} {54, 1341}

\bibitem[\protect\citeauthoryear{{Heard} \& {Gaskell}}{{Heard} \&
  {Gaskell}}{2016}]{Heard+Gaskell16}
{Heard} C.~Z.~P.,  {Gaskell} C.~M.,  2016, \mn@doi [\mnras]
  {10.1093/mnras/stw1616}, \href
  {http://adsabs.harvard.edu/abs/2016MNRAS.461.4227H} {461, 4227}

\bibitem[\protect\citeauthoryear{{Ivanov}, {Polnarev}  \& {Saha}}{{Ivanov}
  et~al.}{2005}]{Ivanov2005}
{Ivanov} P.~B.,  {Polnarev} A.~G.,   {Saha} P.,  2005, \mn@doi [\mnras]
  {10.1111/j.1365-2966.2005.08843.x}, \href
  {http://adsabs.harvard.edu/abs/2005MNRAS.358.1361I} {358, 1361}

\bibitem[\protect\citeauthoryear{{Kalberla}, {Burton}, {Hartmann}, {Arnal},
  {Bajaja}, {Morras}  \& {P{\"o}ppel}}{{Kalberla}
  et~al.}{2005}]{2005A&A...440..775K}
{Kalberla} P.~M.~W.,  {Burton} W.~B.,  {Hartmann} D.,  {Arnal} E.~M.,  {Bajaja}
  E.,  {Morras} R.,   {P{\"o}ppel} W.~G.~L.,  2005, \mn@doi [\aap]
  {10.1051/0004-6361:20041864}, \href
  {http://adsabs.harvard.edu/abs/2005A%26A...440..775K} {440, 775}

\bibitem[\protect\citeauthoryear{{Keel}}{{Keel}}{1980}]{Keel80}
{Keel} W.~C.,  1980, \mn@doi [\aj] {10.1086/112662}, \href
  {http://adsabs.harvard.edu/abs/1980AJ.....85..198K} {85, 198}

\bibitem[\protect\citeauthoryear{{Khachikyan} \& {Weedman}}{{Khachikyan} \&
  {Weedman}}{1971}]{Khachikian+Weedman71}
{Khachikyan} {\'E}.~Y.,  {Weedman} D.~W.,  1971, \mn@doi [Astrophysics]
  {10.1007/BF01001021}, \href
  {http://adsabs.harvard.edu/abs/1971Ap......7..231K} {7, 231}

\bibitem[\protect\citeauthoryear{{Kishimoto} et~al.,}{{Kishimoto}
  et~al.}{2013}]{Kishimoto2013}
{Kishimoto} M.,  et~al., 2013, \mn@doi [\apjl] {10.1088/2041-8205/775/2/L36},
  \href {http://adsabs.harvard.edu/abs/2013ApJ...775L..36K} {775, L36}

\bibitem[\protect\citeauthoryear{{Koay}, {Vestergaard}, {Casasola}, {Lawther}
  \& {Peterson}}{{Koay} et~al.}{2016}]{koay2016}
{Koay} J.~Y.,  {Vestergaard} M.,  {Casasola} V.,  {Lawther} D.,   {Peterson}
  B.~M.,  2016, \mn@doi [\mnras] {10.1093/mnras/stv2495}, \href
  {http://adsabs.harvard.edu/abs/2016MNRAS.455.2745K} {455, 2745}

\bibitem[\protect\citeauthoryear{{Koratkar} \& {Gaskell}}{{Koratkar} \&
  {Gaskell}}{1991}]{Koratkar91}
{Koratkar} A.~P.,  {Gaskell} C.~M.,  1991, \mn@doi [\apjs] {10.1086/191547},
  \href {http://adsabs.harvard.edu/abs/1991ApJS...75..719K} {75, 719}

\bibitem[\protect\citeauthoryear{{Korista} \& {Goad}}{{Korista} \&
  {Goad}}{2001}]{Korista2001}
{Korista} K.~T.,  {Goad} M.~R.,  2001, \mn@doi [\apj] {10.1086/320964}, \href
  {http://adsabs.harvard.edu/abs/2001ApJ...553..695K} {553, 695}

\bibitem[\protect\citeauthoryear{{Leggett} et~al.,}{{Leggett}
  et~al.}{2006}]{leggett2006}
{Leggett} S.~K.,  et~al., 2006, \mn@doi [\mnras]
  {10.1111/j.1365-2966.2006.11069.x}, \href
  {http://adsabs.harvard.edu/abs/2006MNRAS.373..781L} {373, 781}

\bibitem[\protect\citeauthoryear{{Lipunov} et~al.,}{{Lipunov}
  et~al.}{2010}]{lipunov2010}
{Lipunov} V.,  et~al., 2010, \mn@doi [Adv. Astron.] {10.1155/2010/349171},
  \href {http://adsabs.harvard.edu/abs/2010AdAst2010E..30L} {2010, 349171}

\bibitem[\protect\citeauthoryear{{Lira}, {Ar{\'e}valo}, {Uttley}, {McHardy}  \&
  {Videla}}{{Lira} et~al.}{2015}]{Lira2015}
{Lira} P.,  {Ar{\'e}valo} P.,  {Uttley} P.,  {McHardy} I.~M.~M.,   {Videla} L.,
   2015, \mn@doi [\mnras] {10.1093/mnras/stv1945}, \href
  {http://adsabs.harvard.edu/abs/2015MNRAS.454..368L} {454, 368}

\bibitem[\protect\citeauthoryear{{Lyutyj}, {Oknyanskij}  \& {Chuvaev}}{{Lyutyj}
  et~al.}{1984}]{lyutyj1984}
{Lyutyj} V.~M.,  {Oknyanskij} V.~L.,   {Chuvaev} K.~K.,  1984, Sov. Astron.
  Lett., \href {http://adsabs.harvard.edu/abs/1984SvAL...10..335L} {10, 335}

\bibitem[\protect\citeauthoryear{{MacLeod} et~al.,}{{MacLeod}
  et~al.}{2016}]{macleod16}
{MacLeod} C.~L.,  et~al., 2016, \mn@doi [\mnras] {10.1093/mnras/stv2997}, \href
  {http://adsabs.harvard.edu/abs/2016MNRAS.457..389M} {457, 389}

\bibitem[\protect\citeauthoryear{{Maoz} et~al.,}{{Maoz} et~al.}{1991}]{Maoz91}
{Maoz} D.,  et~al., 1991, \mn@doi [\apj] {10.1086/169646}, \href
  {http://adsabs.harvard.edu/abs/1991ApJ...367..493M} {367, 493}

\bibitem[\protect\citeauthoryear{{Moran}, {Halpern}  \& {Helfand}}{{Moran}
  et~al.}{1996}]{moran1996}
{Moran} E.~C.,  {Halpern} J.~P.,   {Helfand} D.~J.,  1996, \mn@doi [\apjs]
  {10.1086/192341}, \href {http://adsabs.harvard.edu/abs/1996ApJS..106..341M}
  {106, 341}

\bibitem[\protect\citeauthoryear{{Noda} et~al.,}{{Noda}
  et~al.}{2016}]{Noda2016}
{Noda} H.,  et~al., 2016, \mn@doi [\apj] {10.3847/0004-637X/828/2/78}, \href
  {http://adsabs.harvard.edu/abs/2016ApJ...828...78N} {828, 78}

\bibitem[\protect\citeauthoryear{{Oknyanskii}}{{Oknyanskii}}{1993}]{oknyanskii1993}
{Oknyanskii} V.~L.,  1993, Astron. Lett., \href
  {http://adsabs.harvard.edu/abs/1993AstL...19..416O} {19, 416}

\bibitem[\protect\citeauthoryear{{Oknyanskii}, {Lyutyi}  \&
  {Chuvaev}}{{Oknyanskii} et~al.}{1991}]{oknyanskii1991}
{Oknyanskii} V.~L.,  {Lyutyi} V.~M.,   {Chuvaev} K.~K.,  1991, Sov. Astron.
  Lett., \href {http://adsabs.harvard.edu/abs/1991SvAL...17..100O} {17, 100}

\bibitem[\protect\citeauthoryear{{Oknyanskij} \& {Horne}}{{Oknyanskij} \&
  {Horne}}{2001}]{oknyanskij2001}
{Oknyanskij} V.~L.,  {Horne} K.,  2001, in {Peterson} B.~M.,  {Pogge} R.~W.,
  {Polidan} R.~S.,  eds,  ASP Conf. Ser. Vol. 224, Probing the Physics of
  Active Galactic Nuclei. p.~149

\bibitem[\protect\citeauthoryear{{Oknyanskij} \& {van Groningen}}{{Oknyanskij}
  \& {van Groningen}}{1999}]{oknyanskij1999b}
{Oknyanskij} V.~I.,  {van Groningen} E.,  1999, in {Gaskell} C.~M.,  {Brandt}
  W.~N.,  {Dietrich} M.,  {Dultzin-Hacyan} D.,   {Eracleous} M.,  eds,  ASP
  Conf. Ser. Vol. 175, Structure and Kinematics of Quasar Broad Line Regions.
  p.~79

\bibitem[\protect\citeauthoryear{{Oknyanskij}, {Lyuty}, {Taranova}  \&
  {Shenavrin}}{{Oknyanskij} et~al.}{1999}]{oknyanskij1999}
{Oknyanskij} V.~L.,  {Lyuty} V.~M.,  {Taranova} O.~G.,   {Shenavrin} V.~I.,
  1999, Astron. Lett., \href
  {http://adsabs.harvard.edu/abs/1999AstL...25..483O} {25, 483}

\bibitem[\protect\citeauthoryear{{Oknyanskij}, {Lyuty}, {Taranova}, {Koptelova}
   \& {Shenavrin}}{{Oknyanskij} et~al.}{2008}]{oknyanskij2008}
{Oknyanskij} V.~L.,  {Lyuty} V.~M.,  {Taranova} O.~G.,  {Koptelova} E.~A.,
  {Shenavrin} V.~I.,  2008, Odessa Astron. Publ., \href
  {http://adsabs.harvard.edu/abs/2008OAP....21...79O} {21, 79}

\bibitem[\protect\citeauthoryear{{Oknyansky}, {Metlova}, {Taranova},
  {Shenavrin}, {Artamonov}  \& {Gaskell}}{{Oknyansky}
  et~al.}{2014}]{oknyansky2014}
{Oknyansky} V.~L.,  {Metlova} N.~V.,  {Taranova} O.~G.,  {Shenavrin} V.~I.,
  {Artamonov} B.~P.,   {Gaskell} C.~M.,  2014, \mn@doi [Astronomy Letters]
  {10.1134/S1063773714090011}, \href
  {http://adsabs.harvard.edu/abs/2014AstL...40..527O} {40, 527}

\bibitem[\protect\citeauthoryear{{Oknyansky}, {Gaskell}  \&
  {Shimanovskaya}}{{Oknyansky} et~al.}{2015}]{oknyansky2015}
{Oknyansky} V.~L.,  {Gaskell} C.~M.,   {Shimanovskaya} E.~V.,  2015, Odessa
  Astron. Publ., \href {http://adsabs.harvard.edu/abs/2015OAP....28..175O} {28,
  175}

\bibitem[\protect\citeauthoryear{{Oknyansky} et~al.,}{{Oknyansky}
  et~al.}{2016a}]{oknyansky9015}
{Oknyansky} V.~L.,  et~al., 2016a, The Astronomer's Telegram, \href
  {http://adsabs.harvard.edu/abs/2016ATel.9015....1O} {9015}

\bibitem[\protect\citeauthoryear{{Oknyansky}, {Huseynov}, {Lipunov},
  {Gorbovskoy}, {Kuznetsov}, {Balanutza}, {Metlov}  \& {Gaskell}}{{Oknyansky}
  et~al.}{2016b}]{oknyansky9030}
{Oknyansky} V.~L.,  {Huseynov} N.~A.,  {Lipunov} V.~M.,  {Gorbovskoy} E.~S.,
  {Kuznetsov} A.~S.,  {Balanutza} P.~V.,  {Metlov} V.~I.,   {Gaskell} C.~M.,
  2016b, The Astronomer's Telegram, \href
  {http://adsabs.harvard.edu/abs/2016ATel.9030....1O} {9030}

\bibitem[\protect\citeauthoryear{{Oknyansky} et~al.,}{{Oknyansky}
  et~al.}{2016c}]{oknyansky9050}
{Oknyansky} V.~L.,  et~al., 2016c, The Astronomer's Telegram, \href
  {http://adsabs.harvard.edu/abs/2016ATel.9050....1O} {9050}

\bibitem[\protect\citeauthoryear{{Osterbrock}}{{Osterbrock}}{1981}]{Osterbrock81}
{Osterbrock} D.~E.,  1981, \mn@doi [\apj] {10.1086/159306}, \href
  {http://adsabs.harvard.edu/abs/1981ApJ...249..462O} {249, 462}

\bibitem[\protect\citeauthoryear{{Penston} \& {Perez}}{{Penston} \&
  {Perez}}{1984}]{penston_perez1984}
{Penston} M.~V.,  {Perez} E.,  1984, \mn@doi [\mnras]
  {10.1093/mnras/211.1.33P}, \href
  {http://adsabs.harvard.edu/abs/1984MNRAS.211P..33p} {211, 33P}

\bibitem[\protect\citeauthoryear{{Peterson} et~al.,}{{Peterson}
  et~al.}{2004}]{Peterson2004}
{Peterson} B.~M.,  et~al., 2004, \mn@doi [\apj] {10.1086/423269}, \href
  {http://adsabs.harvard.edu/abs/2004ApJ...613..682P} {613, 682}

\bibitem[\protect\citeauthoryear{{Pronik} \& {Chuvaev}}{{Pronik} \&
  {Chuvaev}}{1972}]{pronik_chuvaev1972}
{Pronik} V.~I.,  {Chuvaev} K.~K.,  1972, \mn@doi [Astrophysics]
  {10.1007/BF01002159}, \href
  {http://adsabs.harvard.edu/abs/1972Ap......8..112P} {8, 112}

\bibitem[\protect\citeauthoryear{{Rowan-Robinson}}{{Rowan-Robinson}}{1995}]{Rowan-Robinson1995}
{Rowan-Robinson} M.,  1995, \mn@doi [\mnras] {10.1093/mnras/272.4.737}, \href
  {http://adsabs.harvard.edu/abs/1995MNRAS.272..737R} {272, 737}

\bibitem[\protect\citeauthoryear{{Runco} et~al.,}{{Runco}
  et~al.}{2016}]{runco16}
{Runco} J.~N.,  et~al., 2016, \mn@doi [\apj] {10.3847/0004-637X/821/1/33},
  \href {http://adsabs.harvard.edu/abs/2016ApJ...821...33R} {821, 33}

\bibitem[\protect\citeauthoryear{{Shappee} et~al.,}{{Shappee}
  et~al.}{2013}]{shappee2013}
{Shappee} B.~J.,  et~al., 2013, The Astronomer's Telegram, \href
  {http://adsabs.harvard.edu/abs/2013ATel.5010....1S} {5010}

\bibitem[\protect\citeauthoryear{{Shappee} et~al.,}{{Shappee}
  et~al.}{2014}]{shappee2014}
{Shappee} B.~J.,  et~al., 2014, \mn@doi [\apj] {10.1088/0004-637X/788/1/48},
  \href {http://adsabs.harvard.edu/abs/2014ApJ...788...48S} {788, 48}

\bibitem[\protect\citeauthoryear{{Sitko}, {Sitko}, {Siemiginowska}  \&
  {Szczerba}}{{Sitko} et~al.}{1993}]{Sitko1993}
{Sitko} M.~L.,  {Sitko} A.~K.,  {Siemiginowska} A.,   {Szczerba} R.,  1993,
  \mn@doi [\apj] {10.1086/172649}, \href
  {http://adsabs.harvard.edu/abs/1993ApJ...409..139S} {409, 139}

\bibitem[\protect\citeauthoryear{{Suganuma} et~al.,}{{Suganuma}
  et~al.}{2006}]{Suganuma2006}
{Suganuma} M.,  et~al., 2006, \mn@doi [\apj] {10.1086/499326}, \href
  {http://adsabs.harvard.edu/abs/2006ApJ...639...46S} {639, 46}

\bibitem[\protect\citeauthoryear{{Turner} \& {Miller}}{{Turner} \&
  {Miller}}{2009}]{Turner+Miller09}
{Turner} T.~J.,  {Miller} L.,  2009, \mn@doi [\aapr]
  {10.1007/s00159-009-0017-1}, \href
  {http://adsabs.harvard.edu/abs/2009A%26ARv..17...47T} {17, 47}

\bibitem[\protect\citeauthoryear{{Zheng}, {Veilleux}  \& {Grandi}}{{Zheng}
  et~al.}{1991}]{zheng1991}
{Zheng} W.,  {Veilleux} S.,   {Grandi} S.~A.,  1991, \mn@doi [\apj]
  {10.1086/170664}, \href {http://adsabs.harvard.edu/abs/1991ApJ...381..418Z}
  {381, 418}

\bibitem[\protect\citeauthoryear{{Zu}, {Kochanek}  \& {Peterson}}{{Zu}
  et~al.}{2011}]{zu2011}
{Zu} Y.,  {Kochanek} C.~S.,   {Peterson} B.~M.,  2011, \mn@doi [\apj]
  {10.1088/0004-637X/735/2/80}, \href
  {http://adsabs.harvard.edu/abs/2011ApJ...735...80Z} {735, 80}

\bibitem[\protect\citeauthoryear{{Zu}, {Kochanek}, {Koz{\l}owski}  \&
  {Udalski}}{{Zu} et~al.}{2013}]{zu2013}
{Zu} Y.,  {Kochanek} C.~S.,  {Koz{\l}owski} S.,   {Udalski} A.,  2013, \mn@doi
  [\apj] {10.1088/0004-637X/765/2/106}, \href
  {http://adsabs.harvard.edu/abs/2013ApJ...765..106Z} {765, 106}

\makeatother
\end{thebibliography}

\bsp	
\label{lastpage}
\end{document}